\documentclass[preprint]{aastex}

\slugcomment{Published in {\it Scientia}, Vol 2, Spring 2011}

\shorttitle{Transit Timing in WASP-3b}
\shortauthors{Littlefield}

\begin{document}


\title{An Examination of Possible Gravitational Perturbations in the Transit
Timing Variations of Exoplanet WASP-3b}


\author{Colin Littlefield}
\affil{University of Notre Dame,
    Notre Dame, IN 46556}







\begin{abstract}
\citet{macie10} claim to have detected a possible sinusoidal
variation in the transit timing variations of exoplanet WASP-3b, which is currently
the only known planet orbiting the star WASP-3. According to Maciejewski's 
analysis, this signal might be the consequence of gravitational perturbations
caused by a hypothetical second exoplanet in the WASP-3 system. I report five
transit timing measurements from the summer of 2010 which provide
modest support for Maciejewski's proposed sinusoidal signal.
\end{abstract}

\section{Introduction}

Exoplanet transits are eclipses which occur when a planet orbiting another star passes directly between that star and the Earth.
Although no current telescope can actually resolve the miniscule silhouette of the planet against its star, it is possible to
detect the small fade produced as the planet blocks some of the light of its star from reaching Earth. Transits for a particular 
exoplanet occur at very regular intervals, and by examining a planet's transit timing variations (TTVs) --- {\it i.e.}, how early or 
late each transit occurs --- it is theoretically possible to detect the effects of gravitational perturbations from other 
exoplanets in the same system.

\citet{macie10} observed six transits of exoplanet WASP-3b and combined them with
eight others from previous publications. They detected a statistically significant sinusoidal signal in the fourteen
TTVs, which spanned a period of over three years. Maciejewski found that the data are consistent with gravitational
perturbations caused by a hypothetical second planet, WASP-3c, with about 15 Earth masses in a 2.02:1 orbital 
resonance with respect to WASP-3b --- very close to a stable 2:1 resonance. They predict that such a planet would be difficult 
to detect via radial velocity measurements because the star's minute oscillations would obfuscate the weak radial 
velocity signature of WASP-3c.

To test Maciejewski's claim, retired professional astronomer Bruce Gary calculated WASP-3b TTVs
from dozens of observations in the Amateur Exoplanet Archive (AXA), his database of exoplanet transit observations by 
amateur astronomers.\footnote{WASP-3 AXA Light Curves, http://brucegary.net/AXA/WASP3/wasp3.htm}  He concluded that
there was little evidence of sinusoidal variation in the amateur TTVs, many 
of which had relatively high uncertainties. Gary's procedure required each observer to perform basic data analysis, 
such as rejection of outlier data and reference star selection, potentially injecting some measure of subjectivity 
into the process. Consequently, even though Gary performed the final analysis, it might not be appropriate to compare 
observations between different amateur observers, and his analysis is not necessarily conclusive.

\section{Observations}

I observed five WASP-3b transits using an 11-inch Schmidt-Cassegrain telescope and a CCD camera at the observatory 
at the Jordan Hall of Science on the University of Notre Dame campus. Before each observing session,
I synchronized the computer's time with Microsoft's Internet
time server. I then took a continuous series of unfiltered images of WASP-3, spanning the length of the eclipse. 
When possible, I also obtained several hours of data before and after the transit in order to better characterize 
systematic errors. For example, the star's magnitude should theoretically remain constant outside of transit, but 
in practice, it will usually vary slightly as a function of the airmass if the target and reference 
stars are different colors. After each observing session, I reduced the data by applying flat fields and dark frames 
before performing aperture photometry to measure the fluxes of WASP-3 and twenty nearby reference stars. To account for 
night-to-night differences in the point-spread function of the stars, I used different aperture sizes for each session in 
an effort to maximize the quality of the photometry.

\begin{deluxetable}{lcc}
\tablewidth{0pt}
\tablecaption{Transit Time Measurements \label{table1}}
\tablehead{
\colhead{Date 2010} & 
\colhead{Mid-Transit} &
\colhead{TTV}   \\
\colhead{(UT)} &
\colhead{(TBD)} &
\colhead{(min)} \\ }
\startdata
Jun 4    &  2455351.6832$\pm 0.0011$ & $\phantom{-}2.4\pm 1.6$ \\
Jul 22   &  2455399.6999$\pm 0.0015$ & $\phantom{-}0.9\pm 2.2$ \\
Aug 2    &  2455410.7802$\pm 0.0013$ & $-0.2\pm 1.8$ \\
Aug 28   &  2455436.6359$\pm 0.0008$ & $-0.1\pm 1.2$ \\
Sep 8    &  2455447.7155$\pm 0.0008$ & $-2.1\pm 1.1$ \\
\enddata
\end{deluxetable}

In order to derive estimates of important transit parameters, especially the time of midtransit, I used an advanced 
spreadsheet designed by Gary for analyzing photometry of exoplanet
transits.\footnote{Bruce Gary's spreadsheets at http://brucegary.net/book\_EOA/xls.htm }
Caltech's NASA Star and Exoplanet Database,
which subsumed Gary's AXA in 2010, has specifically endorsed his procedure, which determines the best-fit model for each
transit through $\chi^2$-square minimization.\footnote{NASA/IPAC/NExScI, Star and Exoplanet Database, Summary Details for Amateur Light Curves, 
http://nsted.ipac.caltech.edu/NStED/docs/datasethelp/AXA.html} 
This process produced estimates of a variety of important transit parameters,
including midtransit time, along with the corresponding uncertainties. To determine
the expected time of midtransit, $T_C$, I
adopted Maciejewski's ephemeris of
$$T_C(E) = 2454605.56000 [BJD] + E \times (1.8468355 {\rm days}) \; . $$
The TTV is simply the difference between the predicted and observed times of midtransit.
Table~1 lists my individual TTV estimates using the above ephemeris.       

Since Maciejewski's ephemeris and subsequent TTV measurements relied upon the Barycentric Julian Date (BJD) reference
frame and Terrestrial Time (TT) time standard, it was necessary to convert my data into a comparable time stamp. 
To accomplish this, I found the midtransit time in Julian Date and, with the assistance of an online conversion utility, 
I converted it to the BJD frame of reference and barycentric dynamical time standard (TBD).\footnote{Jason Eastman's
Barycentric Julian Date Converter, http://astroutils.astronomy.ohio-state.edu/time/utc2bjd.html}  
BJD\_TT and BJD\_TBD use the solar system's barycenter as their frame of reference, and according to Eastman, they vary from each other by no more
than 50 milliseconds, an insignificant difference for TTV analysis \citep{eastman}. These extremely accurate time stamps compensate for
a variety of factors, such as Earth's orbital motion and gravitational perturbations from other planets in the solar system,
which would otherwise taint TTV data due to the finite speed of light.

\begin{figure}[ht!]
\epsscale{.70}
\plotone{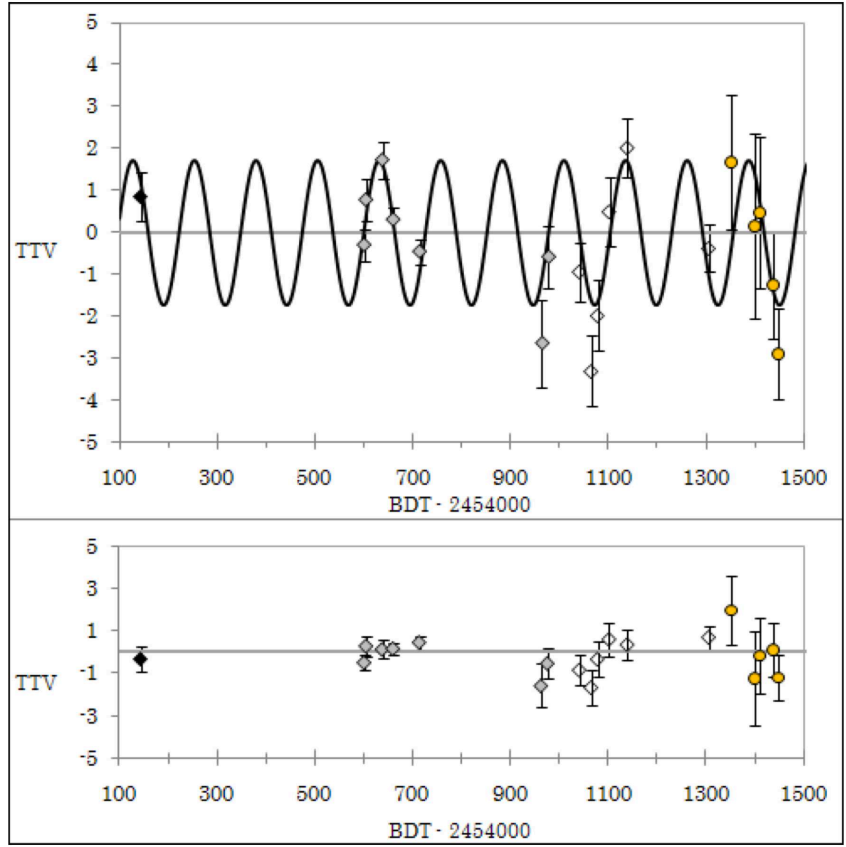}
\caption{The best-fit, single-periodicity model, with TTV estimates superimposed.
The TTV is shown in minutes. The yellow symbols show the new data presented here,
while the other points are from \citet{macie10} and references
therein. The revised ephermeris used in this plot is given in Table~2.
The lower panel shows the residuals to the single-periodicity fit. \label{fig1}}
\end{figure}

\section{Results}

I combined my data with Maciejewski's and fitted it with several different models using $\chi^2$ minimization.
A purely linear model was consistent with Maciejewski's period but the fit was quite poor,
giving a $\chi^2$ parameter of 64.3 with 17 degrees of freedom. 
The resulting p-value of $2.0\times 10^{-7}$ suggests an unacceptable fit to the data. I then fit a three-parameter sinusoid
to the data, retaining the linear term because an inaccurate ephemeris would manifest itself as a linear variation
in the TTV plot. I found that the combined data exhibited a sinusoidal variation with a period of 
125.9$\pm$0.4 days with a full amplitude of 3.4$\pm$0.6 minutes. Furthermore, my analysis calls for a
slightly refined ephemeris for WASP-3b, including a longer orbital period. Table~2 summarizes the parameters of my best-fit model,
including my proposed ephemeris.

Maciejewski proposes both short- and long-term sinusoidal variation in the TTV data, so I performed the analysis again, allowing 
for two independent sinusoids. Without a linear term, the double sinusoid model produced a strong fit; however, once I included 
the requisite linear function to account for possible errors in the ephemeris, the reduced $\chi^2$ parameter
plummeted from a nearly ideal 1.05 ($\chi^2 =13.62$, 13 df) to just 0.5 ($\chi^2 = 5.5$, 11 df), indicating a serious
overfit of the data. The extremely low reduced $\chi^2$ reflects that the total number of parameters in that model
ballooned to eight, an ungainly number for just nineteen observations. Thus, I rejected this model, though I was reluctant
to do so because Maciejewski's computational models predicted that WASP-3c would cause two separate periodicities in WASP-3b's
TTV.

\section{Discussion}

My observations provide equivocal support for Maciejewski's hypothesis. However, five closely-spaced observations with 
a small-aperture telescope are insufficient to provide convincing support for Maciejewski's sinusoidal model, which 
includes high-quality data obtained over a span of over three years. Additional observations would permit a
more rigorous characterization of the TTV of WASP-3b, including an examination of whether there is a second
periodicity in the TTV plot.

One concern is that several of Maciejewski's TTV estimates between BJD 2454900 and 2455100 are somewhat inconsistent
with my fit. The residuals for these observations were much smaller for the double sinusoid model, but 
for reasons already discussed, I decided that a single periodicity best represented the data. Additionally, 
Maciejewski's published TTV estimate for BJD 2455102 is apparently incorrect. I used Maciejewski's ephemeris
and estimated midtransit times to verify the published TTV estimates, and the BJD 2455102 datum showed the
only discrepancy. For that date, Maciejewski's paper lists a TTV of 0.00120 days, but from other data
in the paper I determined that the time of mid-transit corresponded to a TTV of 0.00058 days, a difference
of 0.9 minutes. The most likely explanation is a misprint of that one TTV estimate in his table.

\begin{table}[h!]
\begin{center}
\centerline{Table 2. Best-Fit Model \label{table2}}
\vspace{0.1cm}
\begin{tabular}{lr}
\tableline\tableline
Full Amplitude (min) & 3.7$\pm 0.6$ \\
Sinusoid Period (days) & 125.9$\pm 0.4$ \\
Orbital Period (days) & 1.8468372$\pm 0.0000006$ \\
$T_0$ (BJD)  & 2454605.55978$\pm 0.00010$ \\
$\chi^2$  & 18.21 \\
df  &  14  \\
p-value  & 0.20 \\
\tableline
\end{tabular}
\end{center}
\end{table}

The data quality for my 2010 June~4 observations left much to be desired, potentially making that TTV estimate
unreliable. Jordan Hall is a notoriously vibration-prone building, and that night, high-frequency
vibrations distorted the star images into elongated streaks. In some of the worst images, star images
were over 13 arcseconds long. To put this into perspective, star images at Jordan Hall are normally
circular and no larger than 5 arcseconds in diameter, even on a night of poor seeing. In light of this
problem, it is prudent to be somewhat skeptical of the accuracy of this particular TTV measurement.

On a final note, Maciejewski predicts that the hypothetical second planet would produce a transit
depth of up to 0.35\% if it undergoes transits. Although I did detect a barely significant, 0.3\% fade in the
flux of WASP-3 on 2010 August~1, my own follow-up observations strongly suggest that this feature was spurious,
a product of a common systematic error, such as an imperfect flat-field. My experience suggests that while
the equipment at Jordan Hall is just capable of detecting a transit depth of 0.3\% under ideal conditions,
it is extremely easy for any number of errors to produce false `transits' of that depth.

\section{Conclusion}

Between June and September 2010, I observed five transits of WASP-3b, and my data, when analyzed in conjunction
with previously published observations, suggests that there is indeed significant sinusoidal variation
in WASP-3b's TTVs. Plainly, nineteen combined observations do not constitute a very large sample size, and
more observations are necessary to explore this possibility. Nevertheless, there appears to
be considerable, albeit inconclusive, evidence that a second, undiscovered exoplanet is perturbing
WASP-3b.


\acknowledgments

I wish to thank Professor Peter Garnavich for his guidance with this project. Additionally, my research
was subsidized in part by an AL/SCI UROP grant from Notre Dame's Institute for Scholarship in the Liberal Arts.

\clearpage




\end{document}